# Study of unconfirmed supernovae

L.S. Aramyan[1], A.R. Petrosian[1], A.A. Hakobyan[1], G.A. Mamon[2], D. Kunth[2], M. Turatto[3], V.Zh. Adibekyan[4], T.A. Nazaryan[1]

[1]Byurakan Astrophysical Observatory, Armenia, e-mail: aramyan@bao.sci.am
[2]Institut d'Astrophysique de Paris, France
[3]INAF-Osservatorio Astronomico di Padova, Italy
[4]Centro de Astrofísica da Universidade do Porto, Portugal

*Abstract.* We study the nature of 39 unconfirmed supernovae (SNe) from the sky area covered by Sloan Digital Sky Survey (SDSS) Data Release 8 (DR8), using available photometric and imaging data and intensive literature search. We confirm that 21 objects are real SNe, 2 are Galactic stars, 4 are probable SNe and 12 remain unconfirmed events. The probable types for 4 objects are suggested: 3 SNe are of probable type Ia and SN 1953H is probable type II SN. In addition, we identify the host galaxy of SN 1976N and correct the offsets/coordinates of SNe 1958E, 1972F, and 1976N.

*1. Introduction.* Very recently, new homogeneous data set for 3876 SNe and their 3679 hosts was presented in [1]. In this database, spectroscopic types of several SNe were updated and reclassified. In addition, few objects turned out to be "SN impostors", i.e., eruptions of Luminous Blue Variables (LBVs) [2-4], members of the class of Luminous Red Novae (LRN) [4,5], or Galactic variable stars. In the database there are 39 unconfirmed SNe, which are marked with "?" symbol. Since these SNe are unconfirmed, they are less useful for statistical investigations and excluded from previous studies.

Currently the data set for SNe and their host galaxies presented in [1] is under scrutiny for several statistical studies. To make the unconfirmed SNe useful for these studies, it is important to clarify their nature. For this purpose, we use available photometric and imaging data, and carry out intensive literature search. The obtained results are presented below. The complete study is presented in paper [6].

*2. The sample.* The number of unconfirmed objects in the *total sample* of 3876 SNe is 39 [1]. The primary information of 39 question marked SNe, i.e., coordinates and/or offsets, and their host galaxies comes from [1]. We used the Asiago Supernova Catalogue (ASC) [7] to obtain other necessary information about the unconfirmed SNe (magnitudes, epoch of discovery, discoverers etc.). In all cases the images of host galaxies of these SNe are available from the First Palomar Observatory

Sky Survey (POSS-I), POSS-II, and in a few cases also from the UK Schmidt Telescope (UKST) plates of the southern hemisphere. Since all the SN candidates are from SDSS coverage, we used the SDSS images as a recent imaging data for them. Three SNe candidates (SNe 1991Y, 1992Y, and 2000af) were discovered after 1990, and there is a possibility that additional observations for them exist in the Multi-mission Archive at Space Telescope institute (MAST) and/or European Southern Observatory (ESO) archive. The search of these SNe in the mentioned archives shows no available images.

**3. *The method of analysis.*** We carefully examined the position of SN on available images of the host galaxies and conducted deep literature search. It is important to note that the SNe coordinates and offsets are reported with different levels of accuracy, but fortunately, the precision is generally within 1″. We assumed that the SN is real when the object is visible in two or more images obtained within 400 days from the discovery date, and is not visible in other images taken with more than 400 days from the discovery. For 7 objects with two POSS-I O and E plates available on the same day we did an astrometric comparison of their positions on available images, and considering apparent magnitudes and proper motion limit for an object at a distance ~34 AU due to reflex motion of the Earth [8] we conclude that misidentification of SN as an asteroid is unlikely. We assumed that the object is not a SN and probably it is a projected Galactic star or a SN impostor when it is visible in two or more images, taken more than 400 days apart. In the cases, when only one image for the candidate is available, its stellar like nature was studied and photometry was performed. We counted the object as probable SN if the image did not appear as defect. Finally, we suggested to keep question marks for SNe and count them as unconfirmed SNe for the remaining cases, when even original image of the candidate was absent and no other imaging and useful information was available.

Determination of the probable SN type was performed, according to the morphology of SNe hosts, absolute discovery magnitudes and/or colors of SNe candidates, and SNe position in the host galaxy. Absolute magnitudes at discovery of the objects were calculated and compared with the mean magnitudes of different SNe types [9]. To calculate the absolute magnitudes of SN candidates, we used the method described in [1]. It is well known that core-collapse (CC) SNe avoid early-type galaxies [10], preferably being associated with disk structure [11]. Therefore, if the SN is discovered in early-type galaxy or located far from the disk structure (in case when host galaxy is inclined), we count it as a probable type Ia. Confirmed or probable SNe are classified as CC according to their blue colors and/or absolute magnitudes only.

**4.** *Results.*

**4.1.** *Confirmed SNe.* According to aforementioned criteria, 21 unconfirmed SNe out of 39 are shown to be real SN events. These 21 objects are collected in Table 1, with the names and morphology of their hosts, SNe coordinates, offsets, and magnitudes (photometric band indicated) at the discovery from IAU circulars, as well as with their probable types. A magnitude without band means that the observation has not been made in a standard photometric system (e.g. those reported in the discovery announcement as photographic, blue plate, red plate, CCD without filter, and so on). Anonymous galaxies are listed with the letter "A" followed by the coordinates.

*Table 1.* The list of confirmed objects.

| SN | Galaxy | Morph. | $\alpha_{SN}$ | $\delta_{SN}$ | E/W offset | N/S offset | Discovery *mag* | Probable type |
|---|---|---|---|---|---|---|---|---|
| 1950O | A161509+1857 | Sbc | 16 15 09.3 | +18 57 14.4 | 10.7E | 10.6S | B17 | |
| 1951J | MCG+00-15-01 | SBc | 05 37 52.6 | +00 07 36.9 | 24W | 13N | B17.5 | |
| 1953H | A110342+4951 | S | 11 03 38.7 | +49 50 29.4 | 1E | 3S | B17 | II: |
| 1954ad | UGC4467 | Sb | 08 32 46.3 | +00 13 33.7 | 7.2W | 5.9S | B17.5 | |
| 1955Q | A105606+2409 | Sm | 10 56 08.4 | +24 09 19 | 4W | 7S | B17.5 | |
| 1955R | UGC7740 | Sc | 12 34 42.4 | +49 19 54.8 | 3.8E | 26N | B18 | |
| 1955S | UGC9933 | SBab | 15 36 40.6 | +43 32 38 | 16.2W | 15.8N | B17.5 | |
| 1968J | PGC50284 | S0 | 14 05 52.1 | +53 07 32.3 | 2W | 12S | 16 | Ia |
| 1969G | A123342+0553 | Sb | 12 33 50 | +05 53 42.8 | 4E | 7N | | |
| 1970M | A104818+1403 | S | 10 48 12 | +14 03 15.5 | 13E | 1N | 16.5 | |
| 1972F | MCG+09-20-97 | Sa | 12 07 11.1 | +53 40 32.2 | 17E | 15N | 16 | |
| 1974D | NGC3916 | Sbc | 11 50 55 | +55 09 07.9 | 34E | 31N | 15.5 | |
| 1976A | NGC5004A | SBb | 13 11 01.1 | +29 34 59.9 | 7W | 18N | 16.5 | |
| 1976N | A073200+6513 | S | 07 31 51.4 | +65 12 38.1 | 7E | 6S | 15 | |
| 1980A | MCG+05-29-64A | SBb | 12 20 28.3 | +31 10 10.5 | 9E | 9S | 15.5 | |
| 1980B | MCG+09-19-42 | SBc | 11 19 54.3 | +54 27 46.2 | 16E | 0N | 16 | |
| 1980C | A134524+4745 | | | | 7W | 0N | 17.5 | |
| 1980E | A131942+3414 | Sm | 13 19 42.5 | +34 14 03.9 | 8W | 7S | 16 | |
| 1982X | UGC4778 | Sc | 09 07 01.4 | +50 43 08.7 | 92W | 23N | V16.5 | |
| 1982Y | UGC5449 | SBc | 10 08 00.5 | +68 21 58.7 | 12E | 5N | V17 | |
| 2000af | A114855-0058 | | 11 48 55.7 | –00 58 36.1 | 0E | 0N | R19.8 | |

**4.2.** *Possible Galactic stars.* Out of the 39 unconfirmed SNe two turned out to be with high probability Galactic stars or SN impostors. These two objects are listed in Table 2 where the names and morphology of the hosts, objects coordinates, their offsets, and magnitudes are given too.

*Table 2.* The list of not real SN explosions.

| SN | Galaxy | Morph. | $\alpha_{SN}$ | $\delta_{SN}$ | E/W offset | N/S offset | Discovery *mag* |
|---|---|---|---|---|---|---|---|
| 1954Y | MCG+03-35-37 | SBa | 13 54 30.2 | +15 02 38.7 | 13W | 0N | V19.3 |
| 1973O | NGC7337 | SBb | 22 37 24.5 | +34 21 59.4 | 26W | 28S | V19.0 |

**4.3 *Probable SNe.*** Four objects are suggested to be probable SNe out of the 39 unconfirmed SNe. Table 3 presents information about them and their host galaxies. The last column of Table 3 gives the probable types of two of them.

*Table 3.* The list of probable SNe.

| SN | Galaxy | Morph. | $\alpha_{SN}$ | $\delta_{SN}$ | E/W offset | N/S offset | Discovery *mag* | Probable type |
|---|---|---|---|---|---|---|---|---|
| 1958E | MCG+07-07-72 | SBa | 03 20 40.2 | +42 48 10.2 | 12.2W | 4.4S | 17.5 | |
| 1986P | NGC5763 | S0 | 14 48 58.1 | +12 29 18.3 | 8W | 6S | B17 | Ia |
| 1991Y | A171436+5719 | S0 | 17 14 37.6 | +57 18 26.4 | 0E | 7N | B19 | Ia |
| 1992Y | NGC3527 | SBa | 11 07 19.1 | +28 31 33.8 | 13E | 6S | B18.5 | |

**4.4. *Unconfirmed SNe.*** Because of the lack of information, the nature of 12 objects (SNe 1950M, 1951I, 1955C, 1965O, 1966O, 1968U, 1969Q, 1972T, 1974A, 1984U, 1985K, and 1987R) out of 39 unconfirmed SNe was not clarified and they remain unconfirmed. In all 12 cases, no star-like source is visible in POSS-I, POSS-II, and SDSS images.

*Acknowledgements.* L.S.A., A.R.P., and A.A.H. are supported by the Collaborative Bilateral Research Project of the State Committee of Science (SCS) of the Republic of Armenia and the French Centre National de la Recherché Scientifique (CNRS). This work was made possible in part by a research grant from the Armenian National Science and Education Fund (ANSEF) based in New York, USA. This work was supported by State Committee Science MES RA, in frame of the research project number SCS 13–1C013. V.Zh.A. is supported by grant SFRH/BPD/70574/2010 from FCT (Portugal) and would further like to thank for the support by the ERC under the FP7/EC through a Starting Grant agreement number 239953. M.T. is supported by the INAF PRIN 2011 "Transient Universe".